\documentclass[twocolumn]{aastex701}

\usepackage{amsmath}
\usepackage{CJK}

\newcommand{\swift}{\textit{Swift }}
\newcommand{\swifts}{\textit{Swift}}

\graphicspath{{./}{figures/}}

\begin{document}
\begin{CJK*}{UTF8}{gbsn}
\title{A Disappearing Act:\ Constraints From ``Missing'' Flares of Repeating Partial TDE Candidates}

\author[orcid=0000-0001-9668-2920]{Jason~T.~Hinkle$^{\dagger}$}
\altaffiliation{NHFP Einstein Fellow}
\affiliation{Department of Astronomy, University of Illinois Urbana-Champaign, 1002 West Green Street, Urbana, IL 61801, USA}
\affiliation{NSF-Simons AI Institute for the Sky (SkAI), 172 E. Chestnut St., Chicago, IL 60611, USA}
\email[show]{jhinkle6@illinois.edu}  

\author[orcid=0000-0002-7866-4531]{Chang~Liu (刘畅)$^{\dagger}$}
\affiliation{Department of Physics and Astronomy, Northwestern University, 2145 Sheridan Rd., Evanston, IL 60208, USA}
\affiliation{Center for Interdisciplinary Exploration and Research in Astrophysics (CIERA), Northwestern University, 1800 Sherman Ave., Evanston, IL 60201, USA}
\affiliation{NSF-Simons AI Institute for the Sky (SkAI), 172 E. Chestnut St., Chicago, IL 60611, USA}
\email[]{ptg.cliu@u.northwestern.edu}  

\renewcommand{\thefootnote}{\fnsymbol{footnote}}
\footnotetext[2]{These authors contributed equally to this work.}

\author[orcid=0000-0001-9515-478X]{Adam~A.~Miller}
\affiliation{Department of Physics and Astronomy, Northwestern University, 2145 Sheridan Rd., Evanston, IL 60208, USA}
\affiliation{Center for Interdisciplinary Exploration and Research in Astrophysics (CIERA), Northwestern University, 1800 Sherman Ave., Evanston, IL 60201, USA}
\affiliation{NSF-Simons AI Institute for the Sky (SkAI), 172 E. Chestnut St., Chicago, IL 60611, USA}
\email[]{amiller@northwestern.edu}  

\author[0000-0003-0853-6427]{Ping~Chen}
\affiliation{Institute for Advanced Study in Physics, Zhejiang University, Hangzhou, 310027, China}
\affiliation{Institute for Astronomy, School of Physics, Zhejiang University, Hangzhou, 310027, China}
\email[]{ping.chen@zju.edu.cn}  

\author[orcid=0000-0002-4449-9152]{Katie~Auchettl}
\affiliation{School of Physics, The University of Melbourne, Parkville, VIC, Australia}
\affiliation{Department of Astronomy and Astrophysics, University of California, Santa Cruz, CA, USA}
\email[]{katie.auchettl@unimelb.edu.au}  

\author[0000-0003-4631-1149]{Benjamin~J.~Shappee}
\affiliation{Institute for Astronomy, University of Hawai`i at M\={a}noa, 2680 Woodlawn Dr., Honolulu, HI 96822}
\email[]{shappee@hawaii.edu}  

\author[0000-0001-6017-2961]{Christopher~S.~Kochanek}
\affiliation{Department of Astronomy, The Ohio State University, 140 West 18th Avenue, Columbus, OH 43210, USA}
\affiliation{Center for Cosmology and AstroParticle Physics, The Ohio State University, 191 W.\ Woodruff Ave., Columbus, OH 43210, USA}
\email[]{kochanek.1@osu.edu}


\author[0009-0001-1470-8400]{K. Z. Stanek}
\affiliation{Department of Astronomy, The Ohio State University, 140 West 18th Avenue, Columbus, OH 43210, USA}
\affiliation{Center for Cosmology and AstroParticle Physics, The Ohio State University, 191 W.\ Woodruff Ave., Columbus, OH 43210, USA}
\email[]{stanek.32@osu.edu}  

\author[0000-0003-2377-9574]{Todd~A.~Thompson}
\affiliation{Department of Astronomy, The Ohio State University, 140 West 18th Avenue, Columbus, OH 43210, USA}
\affiliation{Center for Cosmology and AstroParticle Physics, The Ohio State University, 191 W.\ Woodruff Ave., Columbus, OH 43210, USA}
\affiliation{Department of Physics, The Ohio State University, 191 West Woodruff Ave, Columbus, OH 43210, USA}
\email[]{thompson.1847@osu.edu}  

\begin{abstract}

\noindent Recurrent tidal disruption events (rTDEs) are sources that exhibit multiple TDE-like flares; many are likely powered by the recurring partial disruption of a bound star, in a repeating partial TDE (rpTDE). Two such sources, TDE 2022dbl (ASASSN-22ci) and TDE 2020vdq (ZTF20acaazkt), each exhibited two UV/optical flares and, under the assumption of periodicity, both were expected to exhibit a third flare in early 2026. Neither exhibited such a flare, to limits of $L_{\textrm{UV/optical}} \lesssim 10^{42}$ erg s$^{-1}$, $\sim$30$\times$ fainter than the previous flares. Here, we examine several possible explanations. Observing two independent TDEs from the same galaxy within $\sim$2 yr has a probability of $\lesssim$0.5\% for measured average TDE rates and currently expected rate enhancements, unless there is extreme intrinsic dispersion in the rates. Theoretical predictions for a double TDE of both stars in a binary are inconsistent with the observed flares. We therefore conclude that TDE 2022dbl and TDE 2020vdq are rpTDEs. To produce only two observable flares with similar energetics, our semi-analytical modeling strongly favors a main-sequence star promptly placed on a bound orbit with a deep initial tidal encounter at pericenter. These results suggest that the majority of rTDEs with multiple flares over a few-year baseline are likely to be rpTDEs, and that a significant fraction of systems may produce only two observable flares. This has important implications for the use of r(p)TDEs as probes of TDE physics and dynamical processes in the nuclei of other galaxies, in addition to the expected yield from upcoming surveys.

\end{abstract}


\section{Introduction} \label{sec:intro}

\renewcommand{\thefootnote}{\arabic{footnote}}
\setcounter{footnote}{0}

A tidal disruption event (TDE) occurs when an unlucky star is placed on a low angular momentum orbit around a supermassive black hole (SMBH) and, upon encountering the SMBH, gets ripped apart by the intense tidal forces \citep[e.g.,][]{rees88, evans89, phinney89}. TDEs can power luminous accretion flares, with many emitting strongly in the UV/optical bands \citep[see][for a review]{gezari21}. Initial theoretical investigations primarily studied full disruptions \citep{evans89, kochanek94, ulmer99}, where the entire star is unbound in the tidal encounter. Nevertheless, the rate of partial disruptions, where only a fraction of the stellar mass is unbound, is expected to be higher \citep[e.g.,][]{stone20, zhong22}.

In recent years, several TDEs have exhibited multiple UV/optical flares, which we define as recurrent TDEs (rTDEs)\footnote{Several terms have been used to discuss multiple TDE flares from the same galaxy, including ``repeating'' \citep[e.g.,][]{payne21, yao26} and ``recurring'' \citep[e.g.,][]{sun24} TDEs. We advocate for the use of ``recurrent TDE'' when a galaxy is observed to host more than one TDE flare. Adopted from the nova literature \citep[e.g.,][]{chomiuk21}, this separates the observational phenomenon of multiple TDE flares from a particular physical interpretation.}. The short recurrence timescales raise the possibility that we may be witnessing the repeated partial disruption of a single star \citep[e.g.,][]{payne21, wevers23, lin24, yao26}, a so-called repeating partial TDE (rpTDE). The most robust cases, in which more than 3 periodic flares have been detected, are ASASSN-14ko \citep[e.g.,][$>$30 flares]{payne21, payne22, payne23} and AT 2023uqm \citep[][$\gtrsim$5 flares]{wang25}. Interestingly, both of these rpTDEs occur in galaxies that also host AGNs, although the connection to AGN activity, if any, remains unclear. Most rpTDE candidates, however, have only exhibited two flares to date. These include TDE 2018fyk \citep[ASASSN-18ul;][]{wevers19, wevers23, pasham24}, TDE 2019azh \citep[ASASSN-19dj;][]{hinkle21a, yao26}, TDE 2020vdq \citep[ZTF20acaazkt;][]{somalwar25}, TDE 2022dbl \citep[ASASSN-22ci;][]{lin24, hinkle24, makrygianni25}, and TDE 2024pvu \citep[ZTF18abxnvoz;][]{yao26}. There are also several X-ray-selected events: eRASSt J045650.3-203750 \citep{liu23, liu24_eRASSt} and RX J133157.6-324319.7 \citep{malyali23}, and some in AGN host galaxies: AT 2019aalc \citep[ZTF19aaejtoy;][]{veres24}, AT 2021aeuk \citep[ZTF18aanlzzf;][]{sun25}, and AT 2022agi \citep[Gaia22ahd;][in IRAS F01004-2237]{sun24}. The observed properties and flare recurrence times vary widely across the population of events with multiple TDE-like flares.

The most plausible mechanism for producing rpTDEs is the Hills mechanism \citep{hills88, cufari22, lu23}. This occurs when an energetically hard binary passes close to an SMBH and is disrupted, leading to one star being bound to the SMBH on a highly elliptical orbit while the other becomes an unbound hyper-velocity star. This mechanism alone does not produce TDEs. For a TDE to occur and power luminous flares, the pericenter of the bound star must be close to the tidal radius. While the Hills mechanism can place stars on such orbits, two-body scattering and gravitational wave emission likely also play important roles in changing the orbit of the bound star \citep[e.g.,][]{lu23, pan26}. The evolution of rpTDE systems has been studied analytically \citep{Macleod_Spoon_2013, macleod14, Liu_Tidal_2023, bandopadhyay24b}, semi-analytically \citep{broggi24}, and hydrodynamically \citep{bandopadhyay24b, Liu_Repeating_2025}. These results predict that the bound star can undergo several tidal encounters with the SMBH, generally increasing in strength with each pericenter passage, until the star is finally fully disrupted \citep{Liu_Repeating_2025} or ejected from the system \citep{Kiroglu_2023}. Nevertheless, the detailed evolution depends on the stellar structure \citep[e.g.,][]{Macleod_Spoon_2013, bandopadhyay24b}, the stellar density near the SMBH \citep[e.g.,][]{broggi24}, and the initial orbital parameters \citep[e.g.,][]{Liu_Repeating_2025}.

Under the assumption of Hills capture, constraints on the orbital periods of repeated partial TDEs have important implications for the initial binary configuration. In particular, the binary must be hard relative to the central velocity dispersion of a galaxy to survive long enough to be scattered towards the SMBH \citep[e.g.,][]{hills88, cufari22, bandopadhyay24b}. With an estimate of the SMBH mass, the allowed range of initial binary mass and separation can be constrained \citep{cufari22, bandopadhyay24b, hinkle24}. Thus, a population of robust rpTDEs provides unique insight into hard binaries in the nuclear star clusters of other galaxies.

However, the observational signatures of rpTDEs are difficult to accurately predict, given the uncertainties in the mass lost per pericentric passage \citep[e.g.,][]{broggi24, Liu_Repeating_2025} and the poorly constrained radiative efficiency of TDEs \citep[e.g.,][]{svirski17, dai18, mummery25b}. A further complication is that a star initially on a bound orbit may not produce multiple observable flares \citep[e.g.,][]{broggi24, Liu_Repeating_2025}. This can happen if the radius shrinks sufficiently to limit subsequent mass loss, if the star becomes unbound through velocity kicks at pericenter, or if the star is fully disrupted. While observing several cycles is typically required to confirm a (quasi-)periodic source, this may not be possible for some rpTDEs.

In this manuscript, we discuss the two rTDEs, and rpTDE candidates, TDE 2022dbl \citep[ASASSN-22ci;][]{lin24, hinkle24, makrygianni25} and TDE 2020vdq \citep[ZTF20acaazkt;][]{somalwar25}, for which a third flare was not observed at the time expected from extrapolating the temporal separation of the two previous flares. The paper is organized as follows. In Section \ref{sec:missing_flares}, we discuss the long-term observations of TDE 2022dbl and TDE 2020vdq and place constraints on the lack of a third flare. We explore potential explanations for this ``missing'' third flare, as two independent TDEs in Section \ref{sec:independent}, as double TDEs in Section \ref{sec:double}, and as bona fide rpTDEs in Section \ref{sec:rpTDE}. Finally, we discuss our conclusions in Section \ref{sec:disc}. Throughout the paper, we assume a cosmology of $H_0$ = 69.6 km s$^{-1}$ Mpc$^{-1}$, $\Omega_{M} = 0.29$, and $\Omega_{\Lambda} = 0.71$ \citep{wright06, bennett14}.

\section{``Missing'' Flares} \label{sec:missing_flares}

The real-time discovery of TDEs with multiple flares led to the exciting possibility that we could predict when a future flare should occur and observe it, as was done for several flares of ASASSN-14ko \citep{payne22, payne23}. However, as we will discuss in this section, the expected third flares for TDE 2022dbl and TDE 2020vdq did not occur.

\subsection{Previous Observations}

Both TDE 2022dbl \citep[][]{lin24, hinkle24, makrygianni25} and TDE 2020vdq \citep{somalwar25} exhibited two flares with separations of $\approx$2 and $\approx$2.5 yr, respectively. In both cases, the flares were luminous and hot, with timescales and evolution generally consistent with known TDEs. Only the second flare of TDE 2020vdq has spectroscopic coverage, exhibiting the blue continuum and broad H and He lines typical of TDE H+He events \citep{somalwar25}. Both flares of TDE 2022dbl exhibit spectra similar to those of TDE H+He events and are nearly identical in shape after scaling for the flux differences \citep{lin24, hinkle24, makrygianni25}. No significant precursor or earlier flare was observed for either rpTDE candidate \citep[e.g.,][]{hinkle24, somalwar25}.

The strong similarity between the observed flares suggested that these events could be rpTDEs and that a third flare was possible. For TDE 2022dbl, we adopt peak times of MJD = 59635.1 and MJD = 60354.9 for the first and second flares, respectively \citep{hinkle24}. Assuming that the flares should be periodic, a third flare was expected near MJD $\approx$ 61075 (2026 February 04). For TDE 2020vdq, we adopt peak times of MJD = 59113.1 and MJD = 60082.6 for the first and second flares, respectively \citep{somalwar25}. Under the assumption of periodicity, a third flare was expected near MJD $\approx$ 61052 (2026 January 12).

\subsection{Continued Multi-Wavelength Follow-up}

\begin{figure*}[t]
    \centering
    \includegraphics[width=0.98\textwidth]{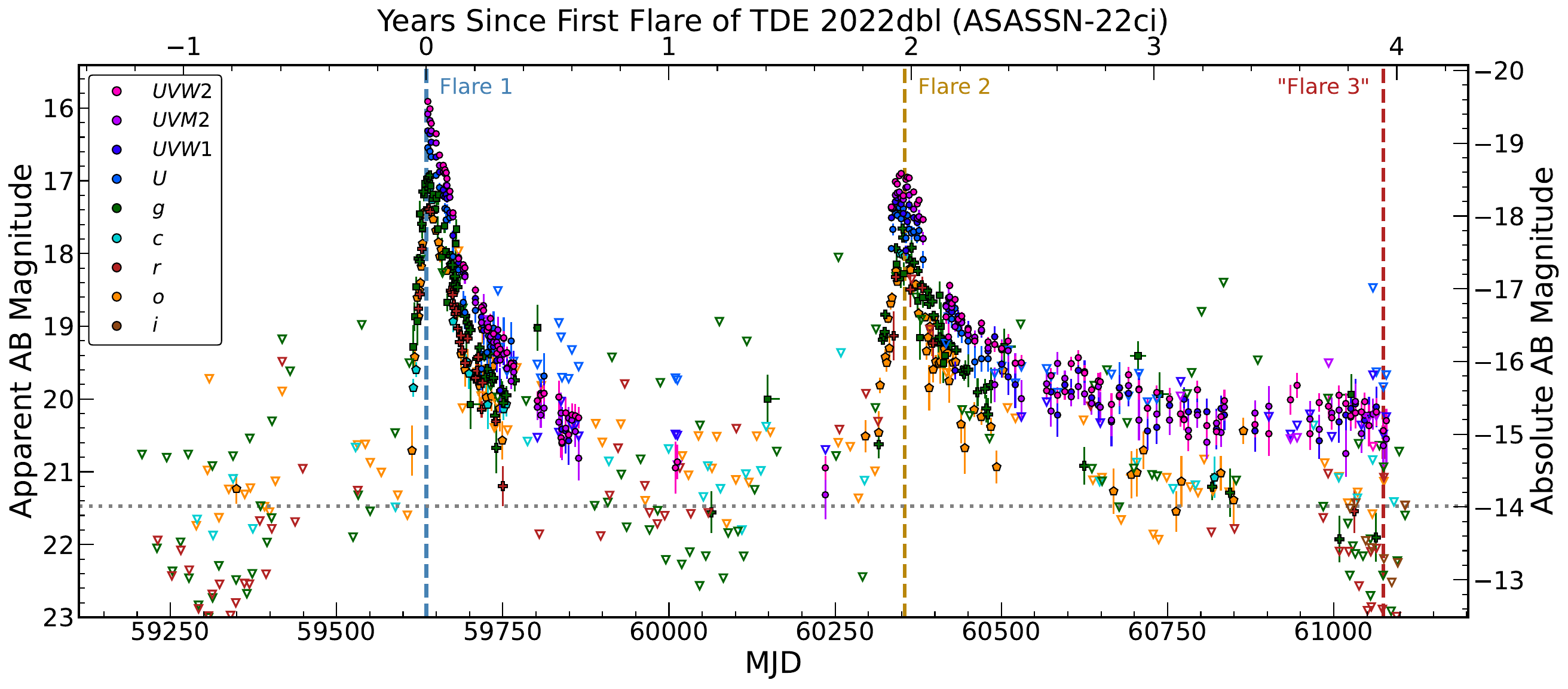}
    \vskip 1em
    \includegraphics[width=0.98\textwidth]{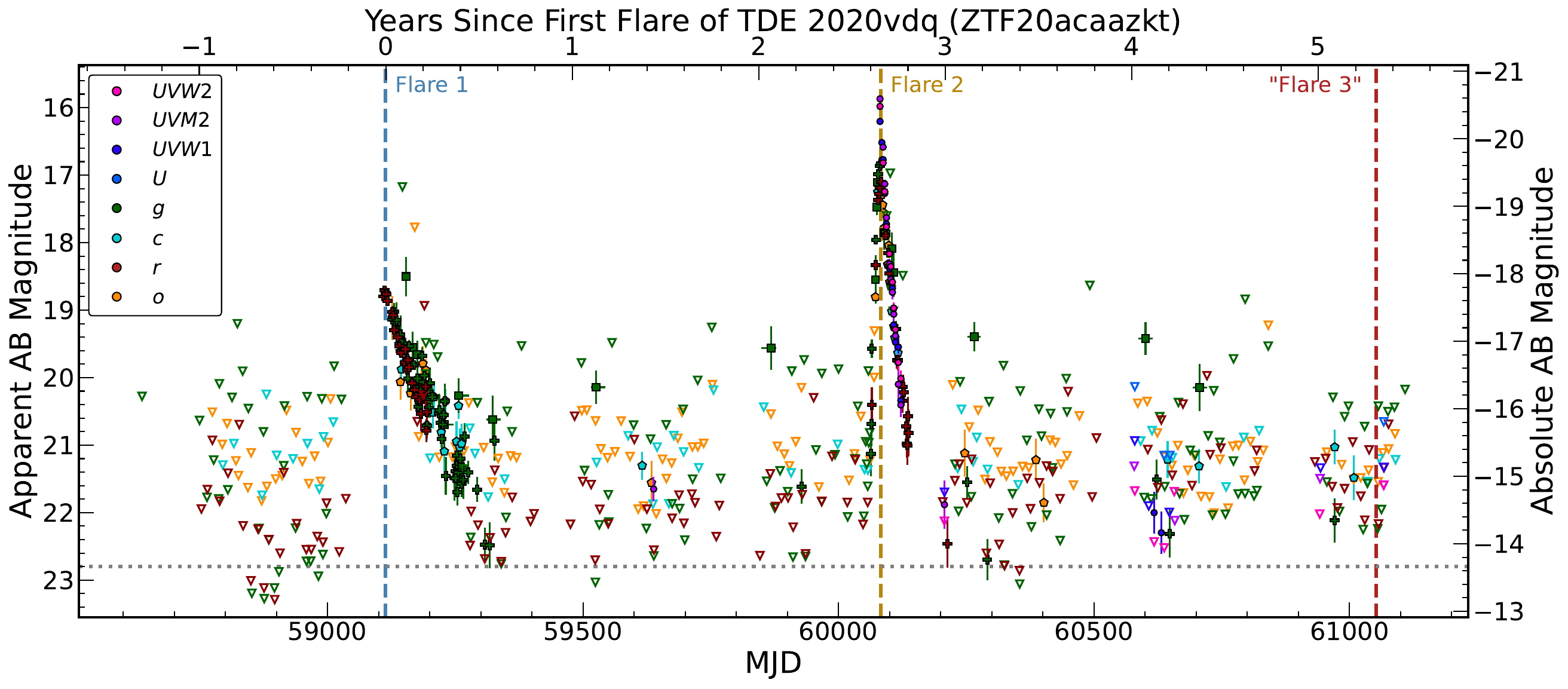}
    \caption{Host-subtracted and Galactic foreground extinction-corrected UV/optical light curves of TDE 2022dbl (ASASSN-22ci; top) and TDE 2020vdq (ZTF20acaazkt; bottom). Data from \swift (UV+$U$; circles), ASAS-SN ($g$; squares), ATLAS ($co$; pentagons), ZTF ($gr$; pluses), and dedicated follow-up photometry of TDE 2022dbl from CDK24 + RC32 ($gri$) are shown. Survey photometry is stacked in 1-d bins during the two flares, except for ASAS-SN data during the first flare of TDE 2020vdq, which is stacked in 10-d bins. To increase the signal-to-noise ratio outside of the flares, we stack the ASAS-SN photometry in 30-d bins and the ATLAS, ZTF, and CDK24 + RC32 photometry in 10-d bins. The \swift data for TDE 2020vdq is stacked in 10-d bins outside of the flares. Downward-facing, open triangles are 3$\sigma$ upper limits for the filter of the same color. The vertical blue and gold dashed lines represent the peak times of the first and second flares, respectively. For each source, the red dashed line marks the expected peak time of the third flare, with UV emission constrained to levels of $M_{\textrm{UV}} \gtrsim -15$. The dotted gray line indicates the approximate level of the UV/optical plateau following the first flare, as measured by \citet{mummery25a} at 3000 \AA. All data are in the AB magnitude system.}
    \label{fig:lightcurves}
\end{figure*}

To monitor for the presence of an additional flare, we continued multi-wavelength follow-up of these sources with the Neil Gehrels Swift Observatory \citep[\swifts;][]{gehrels04} using its UltraViolet and Optical Telescope (UVOT; \citealt{roming05}) and X-ray Telescope (XRT; \citealt{burrows05}). 

\subsubsection{UV/Optical Constraints} \label{sec:uv-opt}

We collected the available epochs of \swift UVOT photometry for TDE 2022dbl (Swift target IDs 15026 and 97566, PIs: Hinkle, Holoien, Margutti, Lin, and Shang) and TDE 2020vdq (Swift target IDs 15030, 97992, and 97620, PIs: Somalwar, Lin, Leloudas, Guolo, Sfaradi, and Hinkle). These observations used the $UVW2$ (2054.6 \AA), $UVM2$ (2246.4 \AA), $UVW1$ (2580.8 \AA), $U$ (3467.1 \AA), $B$ (4349.6 \AA), and $V$ (5425.3 \AA) UVOT filters\footnote{These are the pivot wavelengths from the SVO Filter Profile Service \citep{rodrigo12}.}.

Estimates of the TDE flux require a good estimate of the underlying host galaxy brightness. The host-galaxy UV flux is especially critical for robust measurements of the high-energy TDE flux as it fades. We therefore measured $FUV$ and $NUV$ photometry of the host galaxy from Galaxy Evolution Explorer \citep[GALEX;][]{martin05} data using a 10\farcs{0} radius aperture and \texttt{gPhoton} \citep{million16}. We use the GALEX merged catalog (MCAT) background estimates to more reliably match the GALEX mission photometry, particularly for faint $NUV$ sources \citep[e.g.,][]{million16}. We also obtained host-galaxy magnitudes in the $ugriz$ bands from the Sloan Digital Sky Survey \citep[SDSS;][]{aguado19}, $JHK_S$ bands from the Two Micron All-Sky Survey \citep[2MASS;][]{skrutskie06}, and the $W1,W2$ bands from the Wide-field Infrared Survey Explorer \citep[WISE;][]{wright10} AllWISE survey. For TDE 2022dbl, the optical/IR magnitudes were taken from \citet{hinkle24} and are measured within a 10\farcs{0} radius aperture. For TDE 2020vdq, we use the SDSS cModel magnitudes for $ugriz$ photometry and the AllWISE Source Catalog for $W1,W2$ photometry. The host galaxy of TDE 2020vdq is undetected in 2MASS images.

We used the Fitting and Assessment of Synthetic Templates \citep[\texttt{Fast};][]{kriek09} to fit stellar population synthesis models to the archival host-galaxy photometry. Our fits assume a \citet{cardelli89} extinction law with $\text{R}_{\text{V}} = 3.1$, foreground Galactic extinction as measured by \citet{schlafly11}, a Salpeter IMF \citep{salpeter55}, an exponentially declining star-formation rate, and the \citet{bruzual03} stellar population models. 

We reduced the \swift UVOT data by first summing images within the same observation using the HEASoft \texttt{uvotimsum} package and using \texttt{uvotsource} to measure the photometry. We used a 10\farcs{0} radius aperture for TDE 2022dbl to minimize the effects of the spacecraft pointing jitter present during the second flare. This aperture also includes the entire host-galaxy flux, permitting more straightforward synthetic host-galaxy flux estimation. As there was no significant jitter present during the flares of TDE 2020vdq, we used the default source aperture radius of 5\farcs{0}. In both cases, we used large, nearby, background regions with a radius of $\approx$50\farcs{0}.

To estimate and subtract the host-galaxy flux in the \swift bands, we computed synthetic photometry from the best-fitting \texttt{Fast} host-galaxy SED model. After correcting for foreground Galactic extinction, we additionally fit these \swift UVOT epochs with a blackbody model, as in \citet[][]{holoien21, hinkle22a, hinkle24, hinkle25}.

\subsubsection{X-ray Constraints} \label{sec:x-ray}

We used the epochs taken with the \swift XRT to constrain the X-ray evolution of these sources. Since the detection of X-ray emission between the two flares of TDE 2022dbl \citep[see Figure 4 and Sections 3.5.2 \& 4.5 of][]{hinkle24}, there have been 80 \swift observations of this event, $\sim$30 of which cover the second flare before it faded back to host levels in the optical. To constrain X-ray emission preceding the expected third flare, similar to what was detected between the first and second flares \citep{hinkle24}, we reprocessed all available \swift XRT observations from level one XRT data using the package \texttt{xrtpipeline} version 0.13.7 and applied standard filter and screening criteria with the most recent calibration files. We found no significant ($>$3$\sigma$) X-ray emission associated with individual observations of either flare. To increase the signal-to-noise ratio, we used \texttt{xselect} version 2.5b to merge the individual Swift observations into two groups:\ during the second flare (XL1:\ sw00015026045 -- sw00015026136) and after the second flare had faded (XL2:\ sw00015026080 -- sw00015026136). Neither dataset shows significant X-ray emission, and we derive 3$\sigma$ upper limits of $<1.47 \times 10^{-3}$ count s$^{-1}$ and $<1.73 \times 10^{-3}$ count s$^{-1}$ in the 0.3-10.0 keV energy range for XL1 and XL2, respectively. Assuming an absorbed blackbody with a column density of $N_{H} = 1.94 \times 10^{20}$ cm$^{-2}$ and a temperature of 0.042 keV \citep{hinkle24}, we derive 3$\sigma$ upper limits on the unabsorbed luminosity of $< 1.35\times10^{41}$ erg s$^{-1}$ and $<1.59\times10^{41}$ erg s$^{-1}$ for XL1 and XL2, respectively.

\swift XRT also observed TDE 2020vdq throughout its evolution. We reprocessed all 29 observations that overlapped the position of the event following the same procedures. Using a 150\farcs{0} source-free background region centered at $(\alpha, \delta) = (152.2505184, 42.6086884)$ and a 49\farcs{0} source region centered on the position of TDE 2020vdq, we found no significant ($>$3$\sigma$) X-ray emission except for one individual observation of the second flare of TDE 2020vdq. We found significant ($\sim$3.3$\sigma$) X-ray emission associated with ObsID sw00015030009, which was taken $\sim$20 d (MJD = 60102.23) after the peak of the second flare. To increase the signal-to-noise ratio, we again grouped the \swift observations associated with the second flare of TDE 2020vdq into a bin before the X-ray detection (XU1:\ sw00015030002 -- sw00015030008), the X-ray detection (Xd:\ sw00015030009), after the X-ray detection but still during the UV/optical flare (XU2:\ sw00015030010 -- sw00015030016), and after the UV/optical flare had decayed back to host levels (XU3:\ sw00015030017 -- sw00015030025, sw00097620001 -- sw00097620003, \& sw00097992002 -- sw00097992004). During XU1, XU2 and XU3, we find no significant X-ray emission and derive $3\sigma$ upper limits of $<2.2 \times 10^{-3}$ count s$^{-1}$, $<2.4 \times 10^{-3}$ count s$^{-1}$, and $<1.6 \times 10^{-3}$ count s$^{-1}$ in the 0.3-10.0 keV energy range, respectively. For Xd, we derive a 0.3-10.0 keV count rate of $(4.4 \pm 1.9) \times 10^{-3}$ count s$^{-1}$. Assuming an absorbed blackbody with a column density of $N_{H} = 1.47 \times 10^{20}$ cm$^{-2}$ \citep{HI4PI16} and a temperature of 0.05 keV, similar to TDE 2022dbl, we derive 3$\sigma$ upper limits to the unabsorbed luminosity of $<4.5\times10^{41}$ erg s$^{-1}$, $<4.9\times10^{41}$ erg s$^{-1}$ and $<3.4\times10^{41}$ s$^{-1}$ for XU1, XU1 and XU3, respectively. For Xd, we derive an unabsorbed luminosity of $9.2\pm4.0\times10^{41}$ s$^{-1}$. These are all consistent with the XRT analysis by \citet{somalwar25}.

\subsection{Ground-Based Optical Photometry}

In addition to the \swift data, we obtained public survey data from the All-Sky Automated Survey for Supernovae \citep[ASAS-SN;][]{shappee14, kochanek17}, the Asteroid Terrestrial Impact Last Alert System \citep[ATLAS;][]{tonry18, smith20}, and the Zwicky Transient Facility \citep[ZTF;][]{bellm19, masci19}. For the ASAS-SN $g$ photometry, we rebuilt the reference images for each source using images taken prior to the first flares and performed aperture photometry following the standard ASAS-SN procedures \citep{shappee14, kochanek17}. After removing epochs with poor data quality (e.g., anomalously high PSF widths), we stacked the ASAS-SN data in 1-d bins during the flares and 30-d bins outside of the flares to increase the signal-to-noise ratio. We obtained ATLAS $c$ and $o$ light curves from their forced point-spread function (PSF) photometry service \citep{shingles21} and ZTF $g$ and $r$ photometry from their forced PSF photometry service \citep{masci19}. We stacked the ATLAS and ZTF data in 1-d bins during the flares and 10-d bins outside of the flares.

We additionally obtained dedicated follow-up photometry of TDE 2022dbl in the $gri$ bands with the Corrected Dall-Kirkham 24-inch telescope (CDK24) and Ritchey-Chretien 32-inch telescope (RC32) operated by the Post Observatory from 2025 December 10 to 2026 March 8. The images were taken with Sloan $g$, $r$, and $i$ filters produced by Astrodon. We performed basic image reductions, including bias subtraction, dark subtraction, and flat fielding, with the \texttt{MaxIm DL} version 6.50 software. We built internal reference images by stacking all images observed on 2026 February 19 with a total exposure time of 1500 seconds in each band and used them for image subtraction. Then, we performed forced PSF photometry at the position of TDE 2022dbl. The photometric calibration used SDSS magnitudes transformed from the Pan-STARRS photometric catalog \citep{Flewelling2020_PS1}. These images cover MJD = 61019.5 to MJD = 61107.3, which includes the expected time of the third flare. No trend is observed, so we stacked these data in 10-d bins to obtain deeper limits. Alternatively, stacking all of these images yields deep 3$\sigma$ upper-limits on the presence of a third flare from TDE 2022dbl at $g > 23.6$, $r > 24.1$, and $i > 23.2$ AB mag. Figure \ref{fig:lightcurves} presents the long-term UV/optical photometry of TDE 2022dbl and TDE 2020vdq.

\subsection{Constraints on the Lack of a Third Flare} \label{sec:third_flare_constraints}

\begin{figure*}[t]
    \centering
    \includegraphics[width=0.495\textwidth]{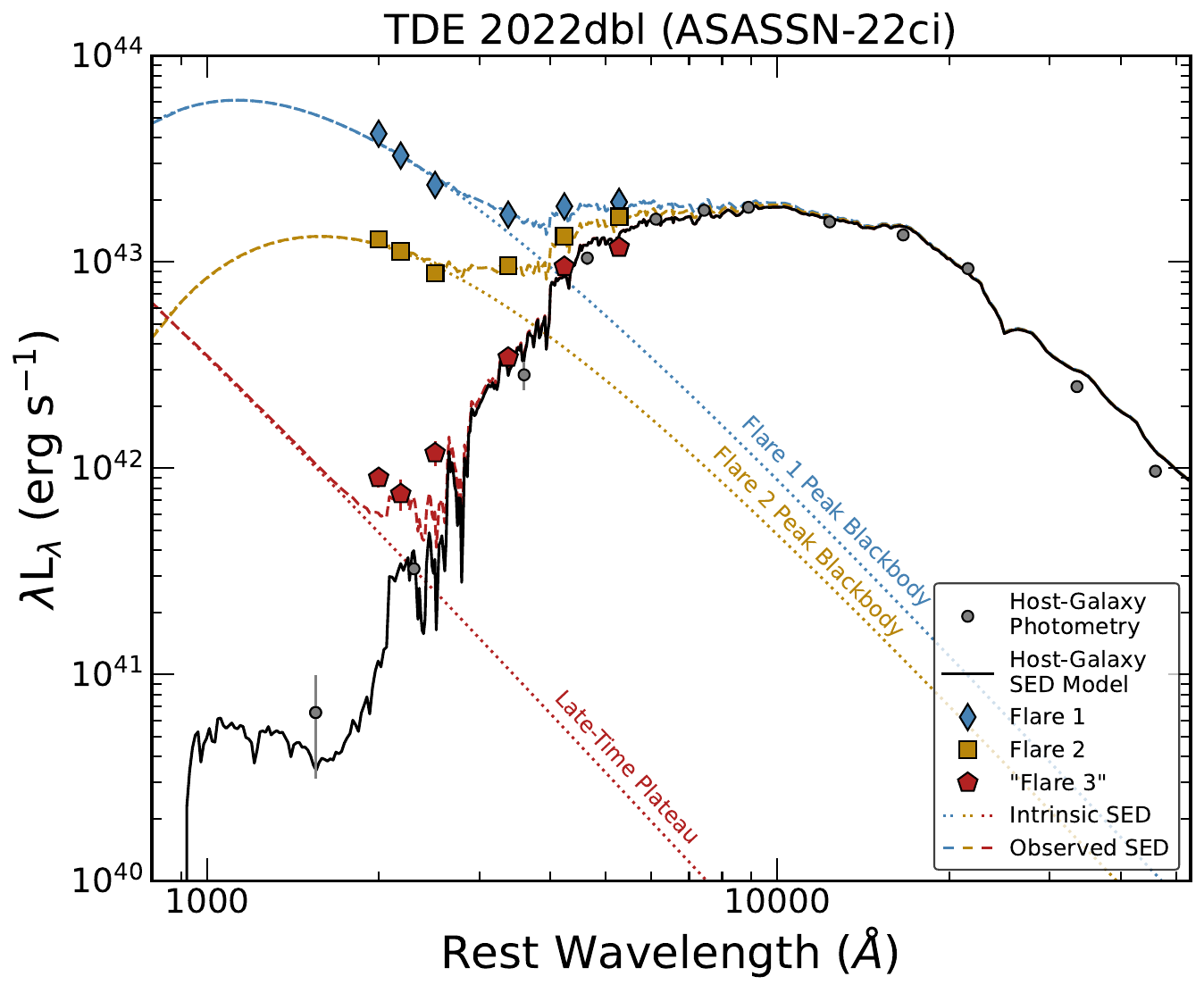}
    \includegraphics[width=0.495\textwidth]{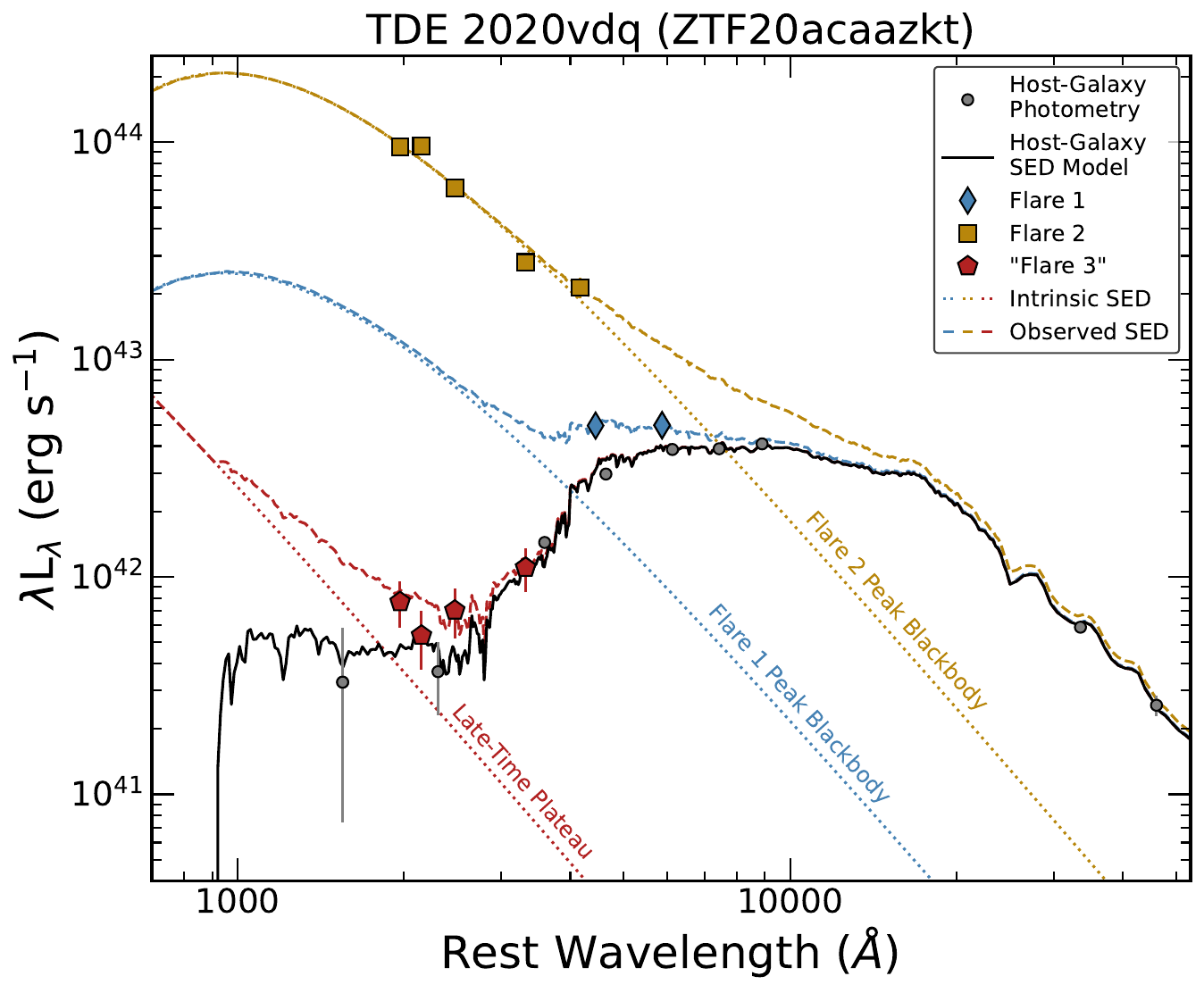}
    \caption{Spectral energy distribution (SED) of the host-galaxy and near the (predicted) peaks for TDE 2022dbl (ASASSN-22ci; left) and TDE 2020vdq (ZTF20acaazkt; right). The host-galaxy photometry is shown in gray circles, and the black line is the best-fitting model. The blue diamonds, gold squares, and red pentagons are the observed SEDs near the time of peak for the first, second, and expected third flares. All photometry and the host-galaxy model have been corrected for Galactic foreground extinction. No host-galaxy light has been removed from the transient photometry. All data have error bars, although they are often smaller than the symbols. The dotted lines correspond to the assumed intrinsic SEDs at the three epochs. For the two observed flares, this is the measured blackbody SED. For the unobserved third flare, this is the expected late-time plateau emission for a temperature of $10^{5.5}$ K \citep[e.g.,][]{guolo25} and scaled to match the $3000$ \AA\ luminosity measured by \citet{mummery25a}. The dashed lines are the observed SEDs after adding the underlying host-galaxy luminosity. The SED at the expected time of the third flare shows no flare and is instead fully consistent with host-galaxy light in the optical, and broadly matches the expectations of a persistent late-time plateau in the UV.}
    \label{fig:multiflare_seds}
\end{figure*}

Recall that the expected peak times of the third flares were MJD $\approx$ 61075 for TDE 2022dbl and MJD $\approx$ 61052 for TDE 2020vdq, indicated with the vertical red lines in Figure \ref{fig:lightcurves}. No transient is observed within one month of the expected peak times.  Given the rise time of $\approx$30 d for TDE 2022dbl \citep{hinkle24} and similar typical TDE rise times \citep{hammerstein23}, we should still expect some flux change even if the peak was later than expected under the simple assumption of strict periodicity (see Section \ref{sec:rpTDE_orbit} for a discussion of deviations from periodicity).

We can directly compute a limit on the luminosity of any third flare from these systems. For TDE 2022dbl, the late-time UV plateau emission is weakly detected in multiple UV filters. We therefore conservatively use the measured blackbody parameters at the expected time of the third flare. This is $L_{\textrm{BB}} = 1.1^{+1.2}_{-0.8} \times 10^{42}$ erg s$^{-1}$. This is broadly consistent with the expected plateau luminosity given the SMBH mass \citep[e.g.,][]{mummery25a}, yet clearly brighter than the plateau following the first flare by $\sim$0.6 mag. Since this is likely plateau emission, any TDE flare must be fainter than this. Even this conservative limit is a factor of $\approx$30 weaker than the fainter second flare of TDE 2022dbl. The UV emission of TDE 2020vdq is not well-detected at the expected time of the third flare, although the UV fluxes are above the estimated host-galaxy levels. We therefore stacked the late-time \swift epochs at MJDs 61065.8 and 61068.6, summed the $\nu L_{\nu}$ luminosities across the $UVW2$, $UVM2$, $UVW1$, and $U$ bands, and multiplied by three to obtain an approximate 3$\sigma$ UV luminosity limit of $L_{UV} \lesssim 1.7 \times 10^{42}$ erg s$^{-1}$. This is also consistent with a late-time plateau and a factor of $\gtrsim$25 fainter than the fainter first flare of TDE 2020vdq.

We can also demonstrate the lack of a third flare by examining the spectral energy distribution (SED) of the TDEs at the relevant epochs. Figure \ref{fig:multiflare_seds} shows the SEDs of TDE 2022dbl and TDE 2020vdq near the peak of their first and second flares and near the time of the expected third flare. While the host-subtracted light curves of Figure \ref{fig:lightcurves} are ideal to elucidate trends in the transient brightness, the absolute levels at faint fluxes are highly sensitive to synthetic photometry estimates of the host-galaxy brightness. To avoid this issue, we present the SEDs without host subtraction, only correcting for Galactic extinction. This demonstrates that the optical emission at the time of the expected third flare is not elevated from that of the host. The UV emission lies slightly above the host-galaxy model and GALEX $NUV$ photometry. This is not evidence of a flare, as these luminosities and temporal evolution are typical of a late-time plateau \citep[e.g.,][]{vanvelzen19a, mummery24, mummery25a}.

Even for phases with detected X-ray emission, the SEDs of TDE 2022dbl and TDE 2020vdq are dominated by UV/optical emission. Furthermore, X-ray emission is not detected from either TDE 2022dbl or TDE 2020vdq at late times, including during the expected third flares. The X-ray non-detections during a phase when the emission is expected to be dominated by direct disk emission \citep[e.g.,][]{guolo25} warrant further study. Possibly, the X-ray disk emission remains obscured by tidal debris and is reprocessed into the UV/optical, which may explain the brighter plateau following the second flare of TDE 2022dbl. Nevertheless, it appears that UV/optical emission is the most reliable tracer of long-term mass accretion for these events, and we therefore focus on the UV/optical observations for the remainder of this paper.

\section{Two Independent TDEs?} \label{sec:independent}

One possibility is that the two flares observed for TDE 2020vdq and TDE 2022dbl are simply independent TDEs. We first examine the probability of such an occurrence given measured TDE rates, the TDE luminosity function, and currently expected rate enhancements in certain types of host galaxies in Section \ref{sec:independent_scaling}. We discuss alternative methods of calculating this probability in Appendix \ref{sec:alternative_stats}, each yielding similar results. However, it is possible that the host galaxies of rTDEs may simply be outliers that the oft-employed tracers of TDE rate enhancements may fail to capture. In Section \ref{sec:dispersion} we estimate the intrinsic dispersion in TDE rates required to plausibly explain the rTDEs as independent TDEs. Section \ref{sec:rates_summary} discusses the implications of these statistical analyses.

\subsection{Scaling from Measured TDE Rates and Enhancements} \label{sec:independent_scaling}

Although finding two independent TDEs within a few years may be unlikely for a typical galaxy, it is well-established that certain types of galaxies, particularly quiescent Balmer-strong and post-starburst galaxies, host TDEs at a rate significantly enhanced relative to an average field galaxy \citep[e.g.,][]{french16, french17, law-smith17, graur18}. We therefore use the measured TDE rate, the TDE luminosity function, and currently expected rate enhancements to estimate the probability of two independent TDEs within a galaxy. A similar calculation was done for TDE 2022dbl and TDE 2020vdq in \citet{makrygianni25}, finding a probability of 0.12\% for TDE 2022dbl and 2.7\% for TDE 2020vdq, assuming the average TDE rate as measured by ZTF \citep{yao23} and a rate enhancement appropriate for each host galaxy. We confirm these results following their assumptions.

However, additional information is present for TDE 2022dbl and TDE 2020vdq, primarily in the form of their peak UV/optical luminosities. Each of the four flares observed for TDE 2022dbl and TDE 2020vdq has a UV/optical luminosity above $10^{43.5}$ erg s$^{-1}$ \citep{hinkle24}\footnote{The first flare of TDE 2020vdq has no UV coverage, limiting the accuracy of a blackbody fit. This luminosity assumes the same temperature as the second flare to scale the optical light curves.}. Given the UV/optical TDE luminosity function, with $\phi(L_{BB}) \propto L_{BB}^{-1.41}$ \citep{yao23}, the rate of TDEs drops dramatically as the peak luminosity of the TDE increases. In fact, with this slope, only 20\% of all TDEs occur with UV/optical luminosities above $10^{43.5}$ erg s$^{-1}$. We note that one flare for each TDE 2022dbl and TDE 2020vdq is even more luminous, at $L_{\textrm{UV/optical}} \gtrsim 10^{44}$ erg s$^{-1}$, which would only decrease the expected rate of similar TDE flares.

Next, we consider the host galaxies in which TDE 2022dbl and TDE 2020vdq reside. The host galaxy of TDE 2022dbl has a Lick H$\delta_A$ index of $2.2 \pm 0.6$ \AA\ and H$\alpha$ emission equivalent width of $0.02 \pm 0.13$ \AA\ \citep{lin24, hinkle24, makrygianni25}. The host-galaxy of TDE 2020vdq has a Lick H$\delta_A$ index of $4.50 \pm 0.13$ \AA\ \citep{makrygianni25} and H$\alpha$ emission equivalent width of $<$1 \AA\ \citep{hinkle24}. The hosts of TDE 2022dbl and 2020vdq are therefore quiescent Balmer-strong and post-starburst, respectively, exactly the galaxies where we expect the TDE rate to be enhanced. We adopt a rate enhancement of 30$\times$, calculated for galaxies with Lick H$\delta_A > 2$\AA\ \citep{french20}, which is therefore appropriate for both hosts.

The rest-frame flare timing separations for TDE 2022dbl and TDE 2020vdq are also similar, at 1.9 yr \citep{makrygianni25, hinkle24} and 2.5 yr \citep{somalwar25}, respectively. We combine the decrement due to the peak luminosities, the rate enhancement, and this flare separation to estimate a rate of TDEs in similar host galaxies over the same timescale. Next, with the 10 known TDEs in galaxies with H$\delta_A > 2$\AA\ and H$\alpha$ emission equivalent width $< 3$\AA\ \citep[e.g.,][]{french20, hinkle24}, we consider the effective number of galaxies searched. Combining this with the Poisson probability for a single galaxy to host two TDEs, we estimate the binomial probability of finding at least one galaxy in this sample with two TDE flares within 2.5 yr at 0.2\%.

\subsection{Intrinsic Dispersion Required to Explain rTDEs as Independent Events} \label{sec:dispersion}

While our analysis above provides useful constraints on the likelihood of finding two TDEs in a given galaxy, it assumes average rates and that the Lick H$\delta_A$ index encodes the necessary information to properly account for TDE rate enhancements. The physics setting TDE rates occur within the SMBH sphere of influence at $\sim$pc scales \citep[e.g.,][]{stone16b, hannah25}, whereas the stellar populations traced by continuum indices are at significantly larger radii. Therefore, we must also address the fact that the galaxies hosting rTDEs may simply be outliers in their TDE rates.

We begin by isolating our analysis to quiescent Balmer-strong and post-starburst galaxies, as their well-established elevated TDE rates make them the most likely candidates as extreme TDE rate outliers. For our previously assumed TDE rate enhancement of $30\times$, the average TDE rate in these galaxies is $9.6 \times 10^{-4}$ galaxy$^{-1}$ yr$^{-1}$. Our fiducial baseline is the 14-yr survey coverage of ASAS-SN. This, combined with the 10 observed TDEs in such systems, sets the effective number of galaxies searched. There are three rTDEs with temporal separations and brightnesses sufficient to have been recovered within this baseline: TDE 2019azh \citep[ASASSN-19dj;][]{hinkle21a}, TDE 2020vdq, and TDE 2022dbl.

First, we assume that the TDE rates for our population of galaxies can be well-described by a Gamma distribution with an average TDE rate fixed at $\mu = 9.6 \times 10^{-4}$ galaxy$^{-1}$ yr$^{-1}$. We then ask what intrinsic dispersion $\sigma$ is required to have a 50\% probability of observing three galaxies with rTDEs having separations of $\leq$14 years.  We adopt a Gamma distribution because it naturally yields a long tail at high TDE rates and the dispersion can be examined analytically. This yields $k = 0.025$, corresponding to $\sigma/\mu = 6.3$, meaning that there must be substantial dispersion to explain these rTDEs as independent TDEs. Even if we assume that one of the three sources with multiple flares in this 14-yr baseline is a genuine rpTDE, this still requires a large dispersion of $\sigma/\mu = 4.9$ for the other flares to be explained as independent events.

We caution that for the right-tailed distributions needed to explain rTDEs as independent events while still obeying the measured average TDE rates, a large intrinsic dispersion reduces the median TDE rate by many orders of magnitude, often to unphysically small values. For the above Gamma distribution with $k = 0.025$, the median TDE rate is $\sim$$3\times10^{-14}$ galaxy$^{-1}$ yr$^{-1}$. Such a scenario is in tension with the predictions of loss cone theory, even after accounting for various processes that may reduce the TDE rate \citep[e.g.,][]{teboul24, polkas24}. Thus, even if rTDEs can be explained as independent events occurring in outlier galaxies with substantially enhanced TDE rates, the distribution of TDE rates cannot be a single smooth distribution.

The assumption that these repeating events are independent TDEs implies very high TDE rates in their host galaxies. Given these elevated TDE rates, we find a 36\% probability of seeing a third independent TDE from these galaxies within a decade. This probability is even higher ($\sim$75\%) if we only consider TDE 2020vdq and TDE 2022dbl, each with two events in 2.5 years. A detection of a third TDE from these galaxies in the near future would provide support for the possibility of rTDEs being independent TDEs. Conversely, the continued non-detection of a third TDE would provide further evidence for the rpTDE scenario, in which the TDE rate need not be significantly enhanced within the host galaxy.

\subsection{Implications for rTDEs} \label{sec:rates_summary}

For estimates starting from measured TDE rates and currently expected rate enhancements, the probability of finding two TDEs within a single galaxy is low, at $\lesssim$0.5\%. We additionally find that explaining the three observed rTDEs with two distinct TDE-like flares with separations shorter than the 14-yr ASAS-SN survey as independent events requires a large intrinsic dispersion within the per-galaxy TDE rates. Although some models have been proposed that can significantly enhance TDE rates for short periods \citep[$\sim$100 Myr; e.g.,][]{stone18, bortolas22, mockler23, melchor24, m_wang24, teboul25}, it is unclear whether they can account for the necessary intrinsic dispersion. A clean observational test of these rTDEs being independent TDEs is that the extremely high implied rates naturally predict a high probability of observing another TDE. Careful long-term archival searches and continued monitoring of rTDE host galaxies will be important in constraining this proposed explanation.

These calculations also indicate that it is generally difficult to explain TDEs exhibiting multiple flares over several-year timescales as independent TDEs. The probability is non-negligible only for events with long flare separations and/or in galaxies with the most extreme rate enhancements. Indeed, the case of TDE 2019azh is instructive. The host galaxy of TDE 2019azh has the most extreme Lick H$\delta_A$ index of any TDE, with estimated rate enhancements reaching as high as $\approx$120$\times$ \citep{french20}. There have been two flares from this host, with a rest-frame separation of 13.2 yr \citep{hinkle21a, yao26}. Following the process discussed in Section \ref{sec:independent_scaling} with the per-galaxy TDE rate from \citet{yao23}, a rate enhancement of 120$\times$, and the fact that this is the sole TDE occurring in a galaxy with such an extreme Lick H$\delta_A$, we find a modest probability of 2.4\% that these two flares were independent TDEs. 

If we assume that none of TDE 2019azh, TDE 2022dbl, nor TDE 2020vdq is truly a rpTDE and a typical TDE rate of $9.6 \times 10^{-4}$ galaxy$^{-1}$ yr$^{-1}$ for similar host galaxies \citep[e.g.,][]{french20}, the probability of all three galaxies exhibiting two independent flares is disfavored at $\sim$4$\sigma$, with only a $4.5 \times 10^{-5}$ probability over the entire ASAS-SN survey baseline. Based on the low probabilities of chance occurrence and the large intrinsic dispersion of the TDE rates required, we conclude that independent TDEs are an unlikely explanation for the two flares observed from TDE 2022dbl and TDE 2020vdq, although it cannot be fully ruled out given the current observations. Nevertheless, the similarity of the two flares of TDE 2022dbl in terms of their spectroscopic and SED properties provides further qualitative support to the rpTDE interpretation rather than them being independent TDEs.

\section{Long-Separation Double TDEs?} \label{sec:double}

Hills capture of one component of a binary within a nuclear star cluster is not the only possible end state of such a system sent on a low angular momentum orbit around an SMBH. It is also possible that both members of the binary pass close enough to the SMBH to be individually disrupted in succession \citep[e.g.,][]{mandel15, mainetti16, bonnerot19, yu24, yu25}. While distinct double-peaked light curves are predicted \citep{mandel15}, other scenarios have theoretical predictions ranging from a precursor feature \citep[][]{bonnerot19}, a knee in the light curve \citep{mainetti16}, to a weaker TDE-like flare \citep{yu25}. This so-called double TDE scenario has been explored in some events with structure in their light curves \citep[e.g.,][]{huang23}, and even invoked to explain some rTDEs, like TDE 2020vdq \cite[e.g.,][]{zhang26}. The remainder of this section discusses why this is unlikely to be a viable channel for producing multiple flares separated by several years.

We begin by noting that only $\approx$18\% of all binaries sent on orbits towards SMBHs result in sequential disruption of both stars \citep{mandel15}. The temporal separation between the disruption of the individual binary components is expected to be small \citep{yu24}. However, unequal mass ratios lead to different return times of the most bound debris \citep{yu24}. Numerical simulations suggest that powering a flare with a time separation of $\Delta T_{\textrm{peak}} / \textrm{min}(t_\textrm{peak}) \gtrsim 1.3$, where $\Delta T_{\textrm{peak}}$ is the temporal separation of the flare peaks and $\textrm{min}(t_\textrm{peak})$ is the shorter rise time, is difficult, with most events having a value below 0.5 \citep{mandel15}. Furthermore, the events with the longest rise times lie at even lower values of $\Delta T_{\textrm{peak}} / \textrm{min}(t_\textrm{peak})$. A fiducial value of 1 and the longest TDE rise times of $\approx$60 d \citep[e.g.,][]{hammerstein23} can only produce peak separations of 60 d. The most extreme $\Delta T_{\textrm{peak}} / \textrm{min}(t_\textrm{peak})$ value of $\sim$2.5 \citep{mandel15} corresponds to a peak separation of $\sim$150 d. Both are much shorter than the observed flare separations for TDE 2022dbl and TDE 2020vdq.

We can also assess the predictions for the mass accretion rate following the disruption of each star. The majority of simulated encounters in \citet{mandel15} result in a ratio of peak accretion rates less than a factor of two, with the highest concentration of simulated outcomes having peak accretion rate ratios within a few tens of percent of unity. The luminosity ratios between the flares of TDE 2022dbl and TDE 2020vdq are approximately a factor of two, suggesting a similar ratio of peak mass accretion rates, in tension with the most common outcomes expected for double TDEs \citep{mandel15}.

There is, however, another channel for producing multiple flares from a double TDE. Approximately $\approx$5\% of binaries sent on orbits towards SMBHs result in the immediate disruption of only one of the stars \citep{mandel15}. Nevertheless, in roughly half of those cases, the surviving binary companion can remain bound to the SMBH with a median orbital period of $\sim$50 years, but possibly as short as 6 months \citep{mandel15}, and can produce another flare at its next pericenter passage. However, this scenario does not naturally account for the extremely similar spectra and flare properties of TDE 2022dbl. Only a small fraction ($<$2.5\%) of double TDEs are expected to lead to flares separated by years, and the median flare recurrence time of $\sim$50 years is much longer than the $\sim$2-yr separations we observe for TDE 2022dbl and TDE 2020vdq. Furthermore, this small fraction conflicts with the observed number and estimated rate of rTDEs \citep[e.g.,][]{yao26}.

Given the low probability of a binary leading to a double TDE and the inconsistency between simulated outcomes and the observations, particularly the temporal separations, we rule out double TDEs as the drivers of the rpTDE candidates TDE 2022dbl and TDE 2020vdq.

\section{Interpretation As Repeating Partial TDEs: Stellar and Orbital Parameters} \label{sec:rpTDE}

\begin{figure*}
    \centering
    \includegraphics[width=\textwidth]{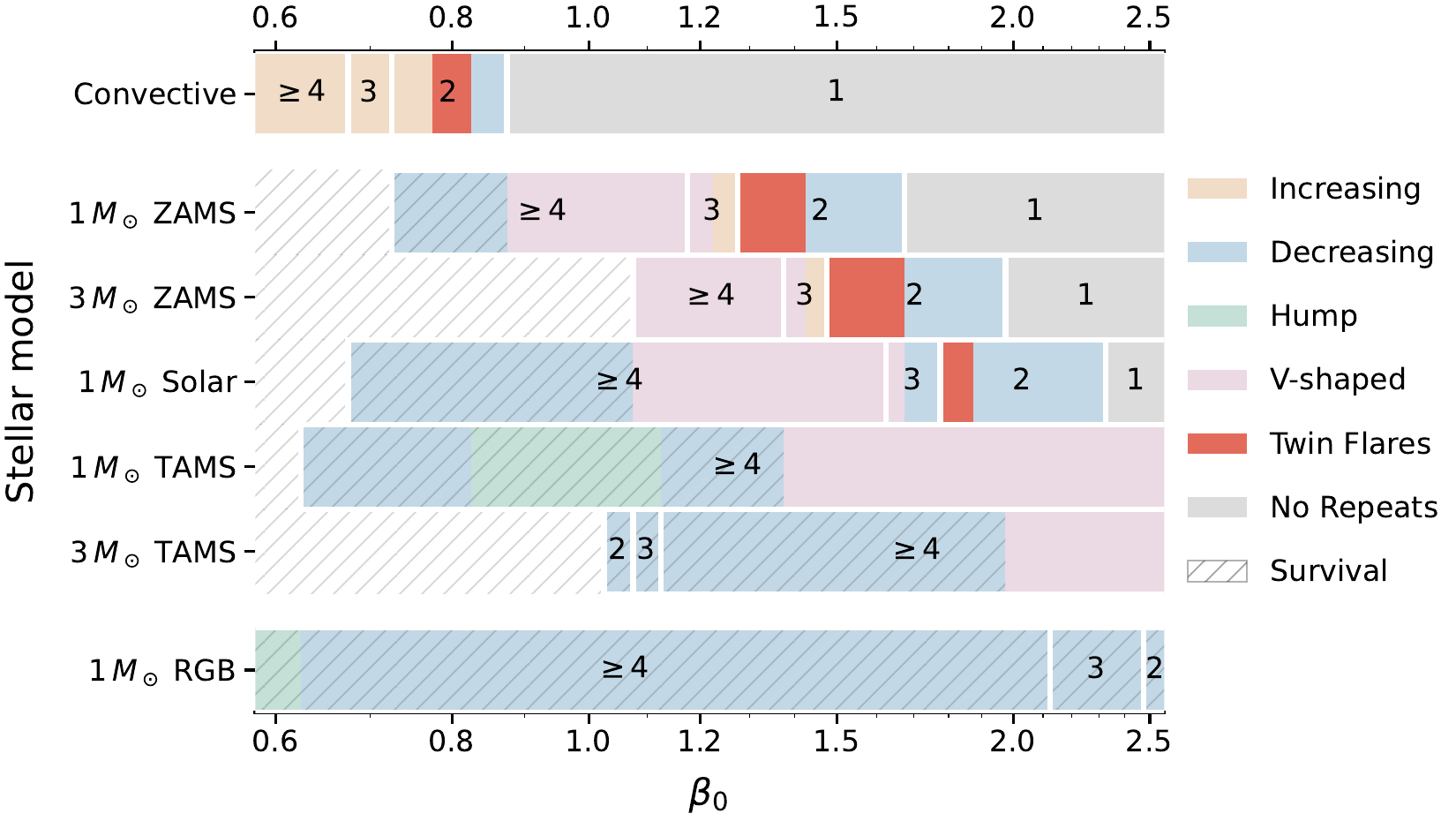}
    \caption{Outcomes of TDE-like encounters for several stellar models across a wide range of initial $\beta_0$. The stellar models include a low-mass fully convective star, 1--3\,$M_\odot$ main-sequence stars at various evolutionary stages (ZAMS, solar-type, and TAMS), and a 1\,$M_\odot$ RGB star. The number of detectable flares, defined by a mass loss threshold of $\Delta M\ge\mathrm{Max}(10^{-3}\,M_\odot, \Delta M_\mathrm{max}/30)$, is indicated in each cell. The background colors represent the underlying trend of the stripped mass across successive pericenter passages (see Section~\ref{sec:rpTDE}). The ``twin flares'' phase space (shown in red) marks regions where exactly two comparable flares are detectable (with a $\Delta M$ ratio within a factor of two), the outcome observed for TDE 2022dbl and TDE 2020vdq. Diagonally hatched regions indicate scenarios where the star ultimately survives its sequence of encounters and the tidal stripping ceases when its radius sufficiently shrinks. Conversely, unhatched regions denote stars that are eventually fully disrupted after multiple passages. The ``no repeats'' class marks stars completely disrupted on their first pericenter passage.}
    \label{fig:DeltaM}
\end{figure*}

The preceding considerations lead us to the natural conclusion that both TDE 2022dbl and TDE 2020vdq were likely rpTDEs that simply did not exhibit a third flare. In this section, we examine scenarios in which a bound star that is initially partially disrupted and returns to pericenter to power a second flare may not get tidally stripped and produce an observable flare on the subsequent passage.

\subsection{Mass Loss Evolution} \label{sec:rpTDE_mass}

To identify the stellar and orbital properties that produce exactly two detectable flares, we test various parameter combinations to determine whether the evolution of the stripped mass ($\Delta M$) over multiple passages can reproduce the observed outbursts. Although converting $\Delta M$ into radiated energy is complex, we reasonably assume that the radiative efficiency remains roughly constant for a given star-SMBH configuration that produces multiple flares. Consequently, the difference in $\Delta M$ should trace the relative luminosity of each flare. 

$\Delta M$ stripped during each pericenter encounter depends on both the stellar structure---particularly the density concentration \citep{guillochon13,Law-Smith_2020,ryu20,Ryu_2020b,Ryu_2020c}---and the instantaneous penetration parameter, $\beta \equiv r_\mathrm{T}/r_\mathrm{p}$, where $r_\mathrm{p}$ is the pericenter distance and the tidal radius is
\begin{equation}
    r_\mathrm{T} = R_*\left(\frac{M_\bullet}{M_*}\right)^{1/3}.
\end{equation}
Any changes in $r_\mathrm{p}$ are negligible \citep{Ryu_2020c,broggi24} compared to the evolution of $r_\mathrm{T}$ as the tidal interaction reshapes the star and alters the stellar radius $R_*$. \citet{Liu_Repeating_2025} demonstrated that even when the star is dramatically deformed following a prior pericenter passage, the $\Delta M$ on the next pericenter passage can still be derived from $\beta$ and the density concentration parameter, $\rho_c/\bar\rho$ (the ratio of central to average density), using the estimates developed for unperturbed, main-sequence star models by \citet{Law-Smith_2020}.

To study the mass stripping processes across different stellar types, we use the Modules for Experiments in Stellar Astrophysics \citep[MESA;][]{Paxton2011, Paxton2013, Paxton2015, Paxton2018, Paxton2019, jermyn23}. We simulate solar-metallicity stars of 1\,$M_\odot$ and 3\,$M_\odot$ at two distinct evolutionary stages: the zero-age main sequence (ZAMS) and the terminal-age main sequence (TAMS). We additionally evolve a 1\,$M_\odot$ model until its central hydrogen fraction matches the current Solar value to mimic a Sun-like star. We also consider a model on the red giant branch (RGB) for a 1\,$M_\odot$ star when the helium core mass reaches 20\% of the ZAMS mass. Finally, to complement the 1\,$M_\odot$ and 3\,$M_\odot$ main-sequence models, which possess extensive radiative zones, we also construct a fully convective, low-mass stellar model using an $n=1.5$ polytrope. 

Following \citet{Macleod_Spoon_2013} and \citet{Liu_Tidal_2023}, we assume that the mass loss proceeds adiabatically. We explicitly neglect tidal heating and stellar spin-up, which would otherwise inflate the star \citep{bandopadhyay24b, Bandopadhyay_2025, Liu_Repeating_2025}. By keeping the specific entropy profile of the MESA model constant as the outer layers are sequentially removed \citep{Dai_2013}, we numerically solve for the star's radius and density profile as a function of its residual mass. During adiabatic mass loss, the polytropic model maintains a fixed $\rho_c/\bar\rho=5.99$ and follows a negative mass-radius relation, $R_{*}\propto M_{*}^{-1/3}$.

The evolution of $\Delta M$ for a star (initial mass $M_{*,0}$, radius $R_{*, 0}$, and penetration parameter $\beta_0$) is calculated in the following iterative process: (i) derive the fractional mass loss $\Delta M/M_*$ from $\beta$ and the stellar properties. For main-sequence stars, we use the fitting formula from \citet{Law-Smith_2020}, where $\Delta M/M$ is a function of $\beta$ and $\rho_c/\bar\rho$. For RGB stars, we use the fitting formula from \citet{Macleod_Spoon_2013}, based on the simulations in \citet{Macleod_Tidal_2012}, where $\Delta M/M_*$ depends on $\beta$, the helium core mass $M_c$, and the polytropic index $n$ of the envelope. $M_c$ remains unchanged throughout the evolution \citep{Macleod_Tidal_2012}, and we fix $n=1.5$ for a fully convective envelope of an evolved star; (ii) update $M_*$; (iii) update $R_*$ and $\rho_c/\bar\rho$ using the results from the adiabatic mass stripping model; (iv) update $\beta$ with the $M_*$ and $R_*$; (v) repeat (i) -- (iv) until the star (or its envelope, if a massive helium core exists) is fully disrupted ($M_* - M_c<10^{-3}\,M_{*, 0}$) or the mass stripping halts and the stellar remnant survives ($\Delta M<10^{-10}\,M_{*, 0}$). 

By scanning over a few stellar types and a wide range of $\beta_0$, we search for the parameter combinations that yield comparable mass loss during the two most significant encounters ($2 > \Delta M_k/\Delta M_{k+1} > 0.5$). To align with the lack of detected precursors in either TDE 2022dbl \citep{hinkle24} or TDE 2020vdq \citep{somalwar25} (also see Section \ref{sec:missing_flares}), any mass loss during prior encounters must be negligible ($\Delta M_{j}/\Delta M_{k} < 1/30$ for $1\le j<k$). Following these two events, the stellar remnant must either undergo complete disruption or contract significantly so that there are no more mass-stripping episodes. Should the star experience any mass loss during subsequent passages, it must be negligible compared to the earlier outbursts ($\Delta M_{j}/\Delta M_{k+1} < 1/30$ for $j>k+1$) to remain consistent with the observations (see Section \ref{sec:third_flare_constraints}).

In Figure~\ref{fig:DeltaM}, we present the outcomes for various stellar types as a function of $\beta_0$. The stellar models in Figure~\ref{fig:DeltaM} are sorted by the $\rho_c/\bar\rho$ ratio, which increases from $\sim$6 for the fully convective star, to $\sim$$10^2$--$10^3$ for the 1--3\,$M_\odot$ main-sequence stars, and to $\sim$$10^7$ for the red giant.
We classify the evolution of $\Delta M$ over successive encounters into four trends: monotonically increasing (orange), monotonically decreasing (blue), a ``hump" profile (initially increasing then decreasing; green), and a V-shaped profile (initially decreasing then increasing; pink). The increasing and V-shaped evolutions exclusively culminate in full disruptions, whereas the decreasing and hump trends generally lead to a surviving stellar remnant. A notable exception occurs at relatively large $\beta_0$ when $\gtrsim$50\% of the stellar mass is stripped in the first pericenter passage; in such cases, the subsequent flare(s) will appear less luminous (as TDE 2022dbl appears to be), mimicking a survival trend, even if the surviving star expands and is ultimately destroyed \citep[see also the hydrodynamical simulations in][]{bandopadhyay24b}. The ``no repeats'' class (gray) marks full disruptions on the stars' first pericenter passage.

Generally, a higher $\beta_0$ yields fewer detectable flares, except for the 3\,$M_\odot$ TAMS model, where the number of detectable flares increases for higher $\beta_0$. This model possesses a deep radiative zone and a developing dense core, and its radius shrinks dramatically when only a small fraction of mass is stripped. For milder encounters ($\beta_0\simeq1.0$), this rapid contraction causes $\Delta M$ to drop quickly below the threshold after just two to three flares, whereas slightly deeper encounters ($\beta_0\gtrsim1.2$) strip enough mass to sustain detectable flaring for many pericenter passages.

The fate of the star is highly dependent on both its internal structure and $\beta_0$. 
Fully convective stars, with a negative mass-radius relation, always expand upon losing mass. Consequently, they become increasingly vulnerable to subsequent tidal stripping, and a complete disruption is inevitable from the onset of the mass loss. 
Conversely, while an RGB star also features a fully convective envelope, its hydrostatic equilibrium is quickly dominated by its massive core once mass loss commences, causing the envelope to shrink \citep[see also][]{Macleod_Spoon_2013, Liu_Tidal_2023}. Within the considered range of $\beta_0\le2.5$, the RGB star consistently survives repeated tidal encounters \citep{Macleod_Tidal_2012}. 
For main-sequence stars with $M_*\gtrsim1\,M_\odot$, the presence of a radiative zone results in a positive mass-radius relation for adiabatic mass loss. 

We identify a critical initial penetration parameter $\beta_\mathrm{surv}$ (the boundary separating the hatched and unhatched regions in Figure~\ref{fig:DeltaM}) as a survival threshold, above which the star will inevitably undergo runaway tidal disruptions. This survivability boundary is heavily dependent on stellar structure. Much like the critical penetration parameter $\beta_\mathrm{crit}$ for full disruption on a single passage \citep{Law-Smith_2020}, $\beta_\mathrm{surv}$ shifts to higher values for more concentrated stars with higher $\rho_c/\bar\rho$, which corresponds to higher ZAMS mass and/or later evolutionary stages. 
Our model overestimates $\beta_\mathrm{surv}$ because it neglects tidal heating effects, which would efficiently inflate the star and drive it toward full disruption at lower $\beta_0$ values. For the same reason, the phase space featuring a V-shaped or decreasing long-term trend could be much narrower, and flares with increasing trends should consequently be more common than expected. This picture might be further complicated by the initial stellar spin following the Hills breakup, where prograde spin could act against the runaway mass loss behavior \citep{Bandopadhyay2026}.

We find that producing two consecutive flares of comparable magnitude (the ``twin flares'' class; red) is only possible when the disrupted object is a main-sequence star lacking a significant core. Most stars, and indeed most stars powering TDEs, are main-sequence stars \citep[e.g.,][]{Macleod_Tidal_2012}. Thus, it is reassuring that systems like TDE 2022dbl and TDE 2020vdq result from main-sequence rpTDEs. In this scenario, the star loses a substantial fraction of its mass ($\geq$1/3) during the first passage, and the remainder is almost entirely stripped during the second. This requires an initial $\beta_0$ slightly below the critical value for a complete disruption, $\beta_\mathrm{crit}$. With only two significant mass-loss episodes, our analysis cannot tightly constrain the initial stellar mass or age as long as $\beta_0$ remains an unconstrained free parameter.
Nevertheless, an evolved giant star progenitor can be safely ruled out. If an RGB star were to lose a significant portion of its envelope during the first encounter ($\beta_0\gtrsim2$), it would contract dramatically, rendering the second flare significantly less luminous than the first; conversely, if the giant lost a much smaller fraction of its envelope, the mass-loss amplitude would evolve far more gradually, resulting in the detection of more than two outbursts \citep{Liu_Tidal_2023, bandopadhyay24b}, which could potentially explain the dozens of flares observed for ASASSN-14ko \citep{payne21, Huang_2023}.

\subsection{Orbital Evolution} \label{sec:rpTDE_orbit}

We also caution that our current framework neglects several physical processes that affect the long-term orbital evolution of the surviving star. We do not account for velocity kicks imparted during the tidal encounter \citep{Manukian_2013,Gafton_kick_2015,Chen_rpTDE_2024} or delayed core reformation \citep{Coughlin_2025}, gravitational wave emission \citep{Peters_1964,Linial_2023b}, dissipation of dynamical tides \citep{Press_1977}, or the hydrodynamical drags in star--disk interaction \citep{Yao_star_disk_2025}. 
In particular, positive-energy kicks during partial disruptions could completely unbind the star from the SMBH \citep{broggi24, Chen_rpTDE_2024}. Typically, the positive orbital energy kick imparted to the star is comparable to its gravitational binding energy \citep{Gafton_kick_2015, Kremer_2022},
\begin{equation}
    \delta E_\mathrm{kick} = \epsilon_\mathrm{kick}\frac{GM_\star}{R_\star},
\end{equation}
where $\epsilon_\mathrm{kick}$ is of order unity, though it scales to smaller values for mild partial disruptions. The fractional change in orbital energy is therefore
\begin{equation}
    \frac{\delta E_{\mathrm{kick}}}{E_\mathrm{orb}} = 0.04\epsilon_\mathrm{kick}\left(\frac{M_\star/R_\star}{M_\odot/R_\odot}\right)\left(\frac{M_\bullet}{10^6\,M_\odot}\right)^{-2/3}\left(\frac{P}{1\,\mathrm{yr}}\right)^{2/3}.
\end{equation}
For short-period repeaters ($P\lesssim$ a few years), this fraction is at most a few percent. Consequently, the resulting shift in the orbital period for a typical system is 
\begin{equation}
    \frac{\delta P}{P}\simeq \frac{3}{2}\frac{\delta E_{\mathrm{kick}}}{E_\mathrm{orb}}\simeq0.06\epsilon_\mathrm{kick}.
\end{equation}
This impact is generally negligible unless the disruption is nearly complete, when $\epsilon_\mathrm{kick}$ can significantly exceed unity in certain dynamical scenarios \citep[e.g.,][]{Coughlin_2025}.

Orbital period evolution can only be measured for rpTDEs with at least three detected flares. Currently this is limited to two optically selected events, ASASSN-14ko \citep[$\dot P<0$;][]{payne21} and AT 2023uqm \citep[$\dot P\simeq0$;][]{wang25}, and the X-ray selected rpTDE candidate, eRASSt J045650.3-203750 \citep[$\dot P<0$;][]{liu23, liu24_eRASSt}. None of them exhibits a net orbital energy injection.
Nevertheless, for higher-$\beta$ encounters and near complete disruptions, a stronger tidal kick could potentially lead to a positive $\dot P$ or even unbind the system. Consequently, the actual number of repeating flares in these systems may be fewer than our fixed-orbit framework predicts, particularly at high $\beta_0$.

Observationally, distinguishing between a full disruption and a stellar ejection is challenging. 
If a surviving star remains on a bound orbit, its subsequent return to pericenter can abruptly interrupt the fallback stream, causing a sudden shut-off in the flare \citep[as proposed for the emission truncation in TDE 2018fyk;][]{pasham24}.
Conversely, if the star is fully destroyed or ejected and never returns to the pericenter, the final flare probably decays more smoothly. As a result, an ejection masks itself as a full disruption through this smooth final decline.

In summary, the absence of a third flare in both TDE 2022dbl and TDE 2020vdq indicates that the disrupted object was a main-sequence star on a highly eccentric orbit, penetrating deeply within the tidal radius ($r_\mathrm{p} \lesssim r_\mathrm{T}$). Rather than being gradually inflated by grazing encounters during previous pericenter passages \citep{Li_2013,Liu_Repeating_2025}, both stars were likely injected directly into orbits with a large initial penetration parameter. This scenario aligns with recent simulations demonstrating that weak tidal interactions are inefficient at inflating main-sequence stars undergoing mass transfer \citep{Yao_Mass_2025}.

\section{Discussion and Conclusions} \label{sec:disc}

\subsection{Repeating Partial TDE Phenomenology}

We have now observed several sources with multiple TDE-like UV/optical flares. We again caution that there is an important distinction between the observation of multiple TDE flares from the same galaxy (a recurrent TDE) and plausible physical explanations of this phenomenon (i.e., independent TDEs and repeating partial TDEs). As the sample of rTDEs grows, so does the diversity of their observed behaviors. This includes the number of observed flares, the long-term luminosity evolution of the flares, and the flare properties themselves. ASASSN-14ko \citep{payne21, payne22, payne23} and AT 2023uqm \citep{wang25}, currently the only events with $\geq$3 nearly periodic flares, are very likely to be bona fide rpTDEs. Nevertheless, their optical spectra, mid-infrared variability, and flare profiles, which occasionally show multiple peaks, are distinct from standard TDEs \citep[e.g.,][]{vanvelzen20b, vanvelzen21, gezari21, hammerstein23} and indicate more AGN-like environments near the SMBH. Such environments could be powered by sustained low-level mass loss from a bound star \citep[e.g.,][]{wang25} or a typical AGN accretion flow \citep[e.g.,][]{tucker20}. Further study of the multi-wavelength emission of rpTDEs, especially in the X-rays to probe the accretion disk, may shed light on this question.

More rpTDE candidates with two observed flares exist. Several, like TDE 2019azh \citep[ASASSN-19dj;][]{hinkle21a, yao26} and TDE 2024pvu \citep{yao26}, have sufficiently long flare separations that searching for previous flares is difficult, and we must wait for the expected time of a third flare. Additionally, the first flares for several candidates lack spectra and often have poor multi-wavelength coverage. Although their physical origin remains uncertain, statistical analyses disfavor unrelated supernovae \citep{yao26} and support the rpTDE scenario \citep[][also see Section\ \ref{sec:independent} and Appendix \ref{sec:alternative_stats}]{makrygianni25, yao26}. It is also important to note that under the assumption of Hills capture, the requirement that the initial binary be hard relative to the central velocity dispersion constrains the resulting orbital period of the bound star to $\lesssim$25 yr \citep[e.g.,][]{hills88, cufari22, bandopadhyay24b, hinkle24}. Therefore, even in galaxies with average TDE rates, two TDE-like flares on longer timescales are either not rpTDEs or require a mechanism for placing a single star on a more weakly bound orbit than those following Hills capture, such as the classic two-body relaxation \citep{pan26} and the eccentric Kozai-Lidov mechanism \citep{melchor24}.

Perhaps most surprising, given theoretical predictions of longer-term mass loss \citep[e.g.,][]{Liu_Repeating_2025}, are events like those studied in this work, TDE 2022dbl and TDE 2020vdq, with just two TDE-like flares on $\approx$2-yr timescales. TDE 2022dbl in particular has excellent observational coverage of both flares, the spectra and SEDs of which are fully consistent with typical TDEs \citep[e.g.,][]{lin24, hinkle24, makrygianni25}. A third flare from these sources is ruled out at luminosities roughly 30 times lower than the previous flares. It is unlikely to observe two unrelated TDEs within a few years from a given galaxy \citep[][also see Section \ref{sec:independent}]{makrygianni25}. Moreover, a single galaxy with a sufficiently high TDE rate to explain these flares as independent TDEs within $\sim$2 yr ought to exhibit another TDE over the $\approx$20-yr baseline covered by surveys like the Catalina Real-Time Transient Survey \citep[CRTS;][]{drake09}, ASAS-SN, ATLAS, and ZTF. No third flares are seen. The two flares seen for TDE 2022dbl and TDE 2020vdq are also inconsistent with theoretical predictions for double TDEs \citep[e.g.,][also see Section \ref{sec:double}]{mandel15, yu24}. Thus, most TDEs with two flares within a few years are likely to be rpTDEs. We find that the progenitors of such systems are relatively unevolved main-sequence stars with orbital pericenters initially close to the critical impact parameter for full tidal disruption. While events like TDE 2022dbl and TDE 2020vdq serve as interesting case studies, we stress that a wide array of outcomes in terms of number of flares and luminosity evolution are expected (see Figure \ref{fig:DeltaM}).

The continued discovery of rpTDEs may lead to several exciting opportunities. For instance, the ability to anticipate and predict future flares can, in principle, allow for targeted observations of the initial rise \citep[e.g.,][]{payne22, payne23} to address important questions on the physical mechanisms powering the early-time UV/optical emission from TDEs \citep[e.g.,][]{metzger16, guillochon16, dai18, mockler19, thomsen22}. The observation of multiple flares from the same system may also lead to better constraints on stellar and SMBH parameters. In particular, the peak luminosity evolution over multiple outbursts opens a unique window into the internal structure of stars in distant nuclear star clusters \citep{Liu_Tidal_2023, Liu_Repeating_2025, bandopadhyay24b, Bandopadhyay_2025}.

\subsection{Connections to Dynamics in Galaxy Nuclei}

Beyond the utility of rpTDEs as testbeds for TDE physics, there are several interesting connections to dynamical processes in galaxy nuclei. Hills capture, and its role in determining the orbits of bound stars in rpTDEs has been studied in detail \citep[e.g.,][]{cufari22, lu23, bandopadhyay24b}. The discovery of several likely rpTDEs in the existing TDE sample suggests that rpTDEs are a substantial fraction of all TDEs \citep[e.g.,][]{yao26}. Moreover, within the Hills capture paradigm, well-constrained orbital periods and SMBH masses for rpTDEs allow for constraints on the initial binaries \citep[e.g.,][]{hinkle24}. In this way, rpTDEs provide a unique pathway towards understanding the properties of hard binaries in other galactic nuclei.

Other repeating phenomena associated with SMBHs have been discovered and studied in recent years, most notably the class of quasi-periodic eruptions \citep[QPEs; e.g.,][]{miniutti19, giustini20, arcodia21, arcodia24, nicholl24, chakraborty25, hernandezgarcia25}. QPEs are repeated soft X-ray flares from the centers of galaxies, with strong connections to TDEs both through similar host environments \citep{wevers22a, wevers24, gilbert25} and the detection of several QPEs following a known TDE or similar accretion flare \citep[][]{nicholl24, chakraborty25, hernandezgarcia25}. Based on these connections and theoretical considerations, the most promising model for QPEs is currently a star in an extreme mass-ratio inspiral \citep[EMRI; e.g.,][]{amaroseoane07, linial17, amaroseoane18, linial23} interacting with a TDE disk \citep[][]{franchini23, linial23, linial25}.

One notable difference between QPEs and rpTDEs is the orbital eccentricities of the bound stars. For QPEs, the quasi-periodicity, long-short timing intervals, and amplitude evolution are most consistent with a mildly eccentric orbit \citep[$e \approx 0.1$;][]{xian21, franchini23}. For rpTDEs, the longer flare separations and the requirement that the star be tidally stripped at pericenter indicate high eccentricity orbits \citep[$e \approx 0.99$;][]{cufari22, Liu_Tidal_2023, bandopadhyay24b}. This indicates a qualitative distinction in the dynamical pathways leading to QPEs and rpTDEs. The mild eccentricities of QPEs indicate EMRI formation with a larger pericenter and long-lived gravitational wave emission that shrinks and circularizes the orbit \citep{lu23,linial25}. In contrast, rpTDEs with bright flares likely begin exhibiting flares promptly after Hills capture and, as we have shown in Section \ref{sec:rpTDE}, are directly placed on orbits with pericenters nearly at the tidal radius. These lead to different expectations on the gravitational wave emission, with rpTDEs potentially powering weak bursts of gravitational waves at each pericenter passage \citep[e.g.,][]{kobayashi04, berry13, toscani20}. The combined multi-messenger study of EMRIs, QPEs, rpTDEs, and TDEs and their various dynamical formation channels thus probes different regimes of loss cone dynamics \citep[e.g.,][]{frank76, lightman77, shapiro78, cohn78} and may provide a holistic view of dynamical processes in other galactic nuclei.

\subsection{Implications for TDE Rates and Future Searches}

An unavoidable consequence of the substantial population of rpTDEs is the implication that the measured optical TDE rate \citep[e.g.,][]{vanvelzen14, holoien16a, yao23} is overestimated. Because rpTDEs can power multiple observable flares, there are several opportunities to detect them. However, to compare the observed TDE rate to theoretical estimates and therefore connect to the physics of scattering and loss-cone dynamics, we are primarily interested in the rate of placing stars on low-angular-momentum orbits rather than the rate of the flares themselves. The most challenging regime for differentiating between an rpTDE and two independent TDEs is when multiple flares are observed on decade timescales. A recent systematic study of long-separation rTDEs leveraging earlier CRTS data in conjunction with ZTF data suggests that rTDEs with flare separations of $<$20 yr could account for $25\%\mbox{ -- }60\%$ of the observed TDE sample \citep{yao26}. Constraining the fraction of all TDE flares coming from r(p)TDEs requires more observations. It will be particularly important to search for weak encounters with faint flares and long temporal separations to investigate the distribution of rTDEs with decade-plus intervals between their flares.

The next decade will provide ample opportunity for robust observational constraints on the rate and properties of r(p)TDEs. The Legacy Survey of Space and Time \citep[LSST;][]{ivezic08} on the Vera C.\ Rubin Observatory and its deep, $g \approx 24.8$ mag, single-epoch limiting magnitude will enable searches for periodic flaring consistent with rpTDEs and precovery efforts at levels $>$2 mag fainter than existing surveys. There is a large range of parameter space that can produce multiple flares (see Section \ref{sec:rpTDE}), and the detection of $>$2 flares gives significantly stronger constraints on the initial stellar and orbital configuration. The continued operation of surveys such as ASAS-SN, ATLAS, and ZTF will increase the temporal baseline over which we can constrain the occurrence of long-separation rpTDEs.

The launch of space-based UV missions like the Ultraviolet Transient Astronomy Satellite \citep[ULTRASAT;][]{shvartzvald24} will provide a complementary view of rpTDEs. ULTRASAT's NUV coverage, large field-of-view, and $\approx$22.5 mag limiting magnitude are ideal for finding TDEs, with several hundred per year expected at low redshifts in its low-cadence survey \citep{shvartzvald24}. Optical multi-band imaging from existing ground-based surveys and upcoming surveys such as the Large Array Survey Telescope \cite[LAST;][]{ofek23}, the Argus Array \citep{law22}, and La Silla Schmidt Southern Survey \citep[LS4][]{miller25}, when combined with ULTRASAT NUV imaging, will provide a powerful temperature diagnostic for transients discovered in ground-based surveys, a means for efficiently identifying TDEs \citep[e.g.,][]{gomez23, stein24}. 

An important caveat is that we expect that nearly full disruptions should power brighter flares. Therefore, events similar to TDE 2022dbl and TDE 2020vdq may represent a sizable fraction of the observed rpTDE population, even if they are a relatively small fraction of the intrinsic rpTDE rate. Events with lower initial $\beta_0$, potentially producing many more flares and diverse evolutionary trends (see Figure~\ref{fig:DeltaM}), are intrinsically fainter and can be systematically missed in flux-limited surveys \citep[e.g.,][]{Liu_Repeating_2025}. Understanding the range of r(p)TDE phenomena and their contamination of the optical TDE rate will require careful multi-wavelength and temporal searches. Regardless, the discovery and characterization of rpTDEs represent an exciting chance to test TDE physics in new regimes and gain insight into otherwise unobservable dynamical processes in the nuclei of other galaxies. Unfortunately, given the several-year separations of most known rpTDEs, we must wait patiently for these opportunities to present themselves.

\begin{acknowledgments}

We thank Decker French, Nicholas Stone, Fred Rasio, and Alexa Anderson for helpful discussions. We thank Richard Post for assistance in obtaining images of TDE 2022dbl (ASASSN-22ci).

J.T.H. acknowledges support from NASA through the NASA Hubble Fellowship grant HST-HF2-51577.001-A, awarded by STScI. STScI is operated by the Association of Universities for Research in Astronomy, Incorporated, under NASA contract NAS5-26555.
C.L. and~A.A.M.~are supported by DoE award \#\,DE-SC0025599, while A.A.M.~is also supported by Cottrell Scholar Award \#\,CS-CSA-2025-059 from Research Corporation for Science Advancement.
C.S.K and K.Z.S are supported by NSF grants AST-2307385 and AST-2407206. 
P.C. acknowledges support from the Zhejiang Provincial TopLevel Research Support Program. 
Parts of this research were supported by the Australian Research Council Discovery Early Career Researcher Award (DECRA) through project number DE230101069.

We thank the \swift PI, the Observation Duty Scientists, and the science planners for promptly approving and executing our \swift target-of-opportunity observations. We acknowledge the use of public data from the Swift data archive. 

ASAS-SN is funded by Gordon and Betty Moore Foundation grants GBMF5490 and GBMF10501 and the Alfred P. Sloan Foundation grant G-202114192.

This work has made use of data from the Asteroid Terrestrial-impact Last Alert System (ATLAS) project. The Asteroid Terrestrial-impact Last Alert System (ATLAS) project is primarily funded to search for near earth asteroids through NASA grants NN12AR55G, 80NSSC18K0284, and 80NSSC18K1575; byproducts of the NEO search include images and catalogs from the survey area. This work was partially funded by Kepler/K2 grant J1944/80NSSC19K0112 and HST GO-15889, and STFC grants ST/T000198/1 and ST/S006109/1. The ATLAS science products have been made possible through the contributions of the University of Hawai`i Institute for Astronomy, the Queen's University Belfast, the Space Telescope Science Institute, the South African Astronomical Observatory, and The Millennium Institute of Astrophysics (MAS), Chile.

Based in part on observations obtained with the Samuel Oschin 48-inch Telescope at the Palomar Observatory as part of the Zwicky Transient Facility project. ZTF is supported by the National Science Foundation under Grants No. AST-1440341 and No. AST-2034437, and a collaboration including Caltech, IPAC, the Weizmann Institute for Science, the Oskar Klein Center at Stockholm University, the University of Maryland, the University of Washington, Deutsches Elektronen-Synchrotron and Humboldt University, Los Alamos National Laboratories, the TANGO Consortium of Taiwan, the University of Wisconsin at Milwaukee, and Lawrence Berkeley National Laboratories, Trinity College Dublin, and IN2P3, France. Operations are conducted by COO, IPAC, and UW. The ZTF forced-photometry service was funded under the Heising-Simons Foundation grant \#\,12540303 (PI: Graham).

This work is based in part on observations made by ASAS-SN and ATLAS. The authors wish to recognize and acknowledge the very significant cultural role and reverence that the summits of Haleakal\=a and Mauna Loa have always had within the indigenous Hawaiian community.  We are most fortunate to have the opportunity to conduct observations from these mountains.

\end{acknowledgments}

\begin{contribution}

J.T.H.\ co-conceived and oversaw the project, acquired new \swift observations, led the observational analysis, contributed to the physical interpretation, and led the writing of the paper. C.L.\ co-conceived the project, led the theoretical analysis, contributed to the physical interpretation, and contributed to the writing of the paper. A.A.M.\ participated in the statistical analysis and contributed to the physical interpretation. P.C.\ obtained reduced optical follow-up photometry of TDE 2022dbl. K.A.\ led the reduction and analysis of \swift XRT data, and contributed to the writing of the paper. B.J.S.\ ran custom ASAS-SN light curves and contributed to the statistical analysis. C.S.K.\ contributed to the statistical and theoretical analyses. All authors reviewed the manuscript.

\end{contribution}

\facilities{ASAS-SN \citep{shappee14, kochanek17}, ATLAS \citep{tonry18, smith20}, ZTF \citep{bellm19, masci19}, Swift (XRT \citep{burrows05} and UVOT \citep{roming05})}

\software{NumPy \citep{numpy}, Matplotlib \citep{matplotlib}, HEAsoft \citep{heasoft14}, \texttt{gPhoton} \citep{million16}, \texttt{Fast} \citep{kriek09}, emcee \citep{foremanmackey13}, MESA \citep{Paxton2011, Paxton2013, Paxton2015, Paxton2018, Paxton2019, jermyn23}}

\bibliography{refs, refs_cliu}{}
\bibliographystyle{aasjournalv7}

\appendix

\section{Alternative Calculations of The Probability of Recurrent TDEs Being Independent Events} \label{sec:alternative_stats}

In addition to the estimate made in Section \ref{sec:independent_scaling}, here we examine several alternative methods to calculate the probability of rTDEs being independent events, given measured TDE rates and currently expected rate enhancements in certain types of host galaxies. In Section \ref{sec:arb_scaling}, we begin by assuming an arbitrarily high TDE rate as a baseline. In Section \ref{sec:birthday}, we evaluate the probability of chance coincidence for two TDEs in the same galaxy. In Section \ref{sec:running}, we compute the conditional probability of seeing multiple ``second'' TDEs from known TDE host galaxies. Each of these calculations provides a similar result to Section \ref{sec:independent_scaling}, underscoring the low probability of rTDEs being independent TDEs within the same host galaxy.

\subsection{An Arbitrarily High TDE Rate} \label{sec:arb_scaling}

Here we perform a simple calculation of the likelihood of finding two independent TDEs within a single galaxy, assuming an unusually high TDE rate in that galaxy. Let us assume that the TDE rate within some galaxy is enhanced to 1 every 100 years, approximately the supernova rate of a typical TDE host galaxy. Interestingly, this rate is nearly identical to the maximum TDE rate that can be accommodated by considering the average rate of TDEs per galaxy \citep[$3.2 \times 10^{-5}$ galaxy$^{-1}$ yr$^{-1}$;][]{yao23} and the most extreme rate enhancement estimated for post-starburst galaxies \citep[$\approx$305$\times$;][]{french16, french20}.

The assumed rate of $0.01$ TDEs per year in a given galaxy and a generic 10-yr baseline of survey operations \citep[e.g.,][]{shappee14, kochanek17, tonry18, bellm19} yields an expected 0.1 TDEs. With this expectation and Poisson statistics, we find a 0.5\% probability of finding 2 TDEs within a single galaxy.

\subsection{Probability of Chance Coincidence} \label{sec:birthday}

Another alternative approach is inspired by the classic ``birthday problem'' in probability theory \citep[e.g.,][]{feller68}. We assume the probability of observing a TDE in a given galaxy over the survey duration is constant across an ensemble of $N_{\rm gal}$ galaxies. 
This observable probability naturally folds in both the intrinsic TDE rate and the survey's selection effects. If a survey discovers a total of $N_{\rm TDE}$ events within this ensemble, we can evaluate the probability that at least one galaxy hosts two or more independent TDEs.
In practice, TDE rates can be elevated in post-starburst galaxies, and the chance of observing a TDE in a given galaxy is not always uniform. Consequently, the calculations below should be treated as order-of-magnitude estimates.

The probability of having all $N_{\rm TDE}$ events discovered in entirely different galaxies is given by:
\begin{equation}
\begin{aligned}
    P(\mathrm{distinct}) & = \frac{N_{\rm TDE}!}{N_{\rm gal}^{N_{\rm TDE}}} \binom{N_{\rm gal}}{N_{\rm TDE}} \\
    & = \prod_{k=0}^{N_{\rm TDE}-1}\left(1-\frac{k}{N_{\rm gal}}\right).
\end{aligned}
\end{equation}
For $N_{\rm TDE} \ll N_{\rm gal}$, we can apply the approximation $1-x \simeq e^{-x}$ to simplify the product:
\begin{equation}
    P(\mathrm{distinct}) \simeq \prod_{k=0}^{N_{\rm TDE}-1}\exp\left(-\frac{k}{N_{\rm gal}}\right) \simeq \exp\left(-\frac{N_{\rm TDE}^2}{2N_{\rm gal}}\right).
\end{equation}
Therefore, the probability of finding multiple TDEs in at least one galaxy is:
\begin{equation}
    P(\ge\textrm{2 TDEs}) = 1 - P(\mathrm{distinct}) \simeq \frac{N_{\rm TDE}^{2}}{2 N_{\rm gal}} ,\label{eq:repeat_prob}
\end{equation}
which is valid in the regime where $N_{\rm TDE}^2 / 2N_{\rm gal} \ll 1$.

As a baseline, we adopt the calculations of \citet{yao23} for the average TDE population. They used 33 optical TDEs selected by ZTF over 3 yr to calculate an average TDE rate of $3.2 \times 10^{-5}$ galaxy$^{-1}$ yr$^{-1}$. We can estimate an effective number of galaxies searched ($N_{\rm gal}$) by dividing the total number of observed TDEs by this rate and the survey duration. Because the apparent TDE rate per galaxy is strictly lower than the intrinsic rate due to selection effects and survey incompleteness, the actual $N_\mathrm{gal}$ is higher and this calculation hence yields a lower limit for $N_{\rm gal}$ (and hence upper limit for the probability of finding $\ge2$\,TDEs in the same galaxy). 
Using Equation (\ref{eq:repeat_prob}), we find a 0.2\% probability of finding two TDEs in a single galaxy. A similar probability of 0.1\% is found when considering the subsample of 19 ZTF TDEs with detailed host-galaxy information studied by \citet{hammerstein23}.

However, as discussed previously, some galaxies are known to host TDEs at a higher rate. Let us again consider post-starburst galaxies. Two of the TDEs studied by \citet{hammerstein23} reside in post-starburst galaxies with a Lick H$\delta_A$ index of $>$5 \AA. The expected TDE rate enhancement in galaxies with a Lick H$\delta_A$ index $>$5 \AA\ is $\approx$50$\times$ \citep{french20}. This yields a probability of 0.5\% of finding two independent TDEs within a single post-starburst galaxy, despite the large rate enhancements within these galaxies.

\subsection{Expected Number of TDEs Within a Fixed Timespan} \label{sec:running}

Finally, we consider the conditional probability of finding a second TDE in a galaxy that has already been observed to host a TDE. The expected number of TDEs in some window of time is given by:
\begin{equation}
    n = Nr\Delta t,
\end{equation}
for $N$ ``first'' TDEs, TDE rate $r$, and timespan $\Delta t$. Following our fiducial assumptions discussed in Section \ref{sec:dispersion}, we assume $N = 10$, $r = 9.6\times10^{-4}$  galaxy$^{-1}$ yr$^{-1}$ (i.e., the measured average TDE rate with a $30\times$ enhancement), and $\Delta t = 14$ years. This yields an expectation value of $n = 0.134$ TDEs.

If we consider only TDE 2020vdq and TDE 2022dbl, the probability of at least two ``second'' TDEs is given by:
\begin{equation}
    P(\geq2) = 1 - e^{-n}(1+n),
\end{equation}
which for the above n, yields 0.8\%. However, we note that this calculation ignores the declining TDE luminosity function explored in Section \ref{sec:independent_scaling}, which would lower the probability substantially.

Furthermore, to compute a reasonable population-level probability, we must also account for TDE 2019azh. The probability of at least three ``second'' TDEs is given by:
\begin{equation}
    P(\geq3) = 1 - e^{-n}\left(1+n+\frac{n^2}{2}\right),
\end{equation}
which for $n = 0.134$ yields a very small probability of $0.04$\%. This again suggests that it is difficult to explain each pair of flares from TDE 2019azh, TDE 2020vdq, and TDE 2022dbl as independent TDEs.

\end{CJK*}
\end{document}